\documentclass[10pt]{article}

\usepackage{amsmath}
\usepackage{amssymb}

\usepackage{graphicx}

\usepackage{cite}

\usepackage{color}
\usepackage[labelfont=bf,labelsep=period,justification=raggedright]{caption}

\makeatletter
\renewcommand{\@biblabel}[1]{\quad#1.}
\makeatother

\date{}

\begin{document}

\begin{flushleft}
{\Large
\textbf{Cooperation, Norms, and Revolutions: A Unified Game-Theoretical Approach}
}
\\
Dirk Helbing$^{1,2,3,\ast}$,
Anders Johansson$^{1,4}$
\\
\bf{1} ETH Zurich, Zurich, Switzerland
\\
\bf{2} Santa Fe Institute, Santa Fe, New Mexico, USA
\\
\bf{3} Collegium Budapest---Institute for Advanced Study, Budapest,
Hungary
\\
\bf{4} University College London, Centre for Advanced Spatial
Analysis, London, UK
\\
$\ast$ E-mail: Corresponding dhelbing@ethz.ch
\end{flushleft}

\section*{Abstract}
\subsection*{Background}
Cooperation is of utmost importance to society as a whole, but is often challenged by individual self-interests. While game theory has studied this problem extensively, there is little work on interactions within and across groups with different preferences or beliefs. Yet, people from different social or cultural backgrounds often meet and interact. This can yield conflict, since behavior that is considered cooperative by one population might be perceived as non-cooperative from the viewpoint of another.

\subsection*{Methodology and Principal Findings}
To understand the dynamics and outcome of the competitive interactions within and between groups, we study game-dynamical replicator equations for multiple populations with incompatible interests and different power (be this due to different population sizes, material resources, social capital, or other factors). These equations allow us to address various important questions: For example, can cooperation in the prisoner's dilemma be promoted, when two interacting groups have different preferences? Under what conditions can costly punishment, or other mechanisms, foster the evolution of norms? When does cooperation fail, leading to antagonistic behavior, conflict, or even revolutions? And what incentives are needed to reach peaceful agreements between groups with conflicting interests?

\subsection*{Conclusions and Significance}
Our detailed quantitative analysis reveals a large variety of interesting results, which are relevant for society, law and economics, and have implications for the evolution of language and culture as well.

\section*{Introduction}

\begin{figure}[!ht]
\begin{center}
\includegraphics[width=4in]{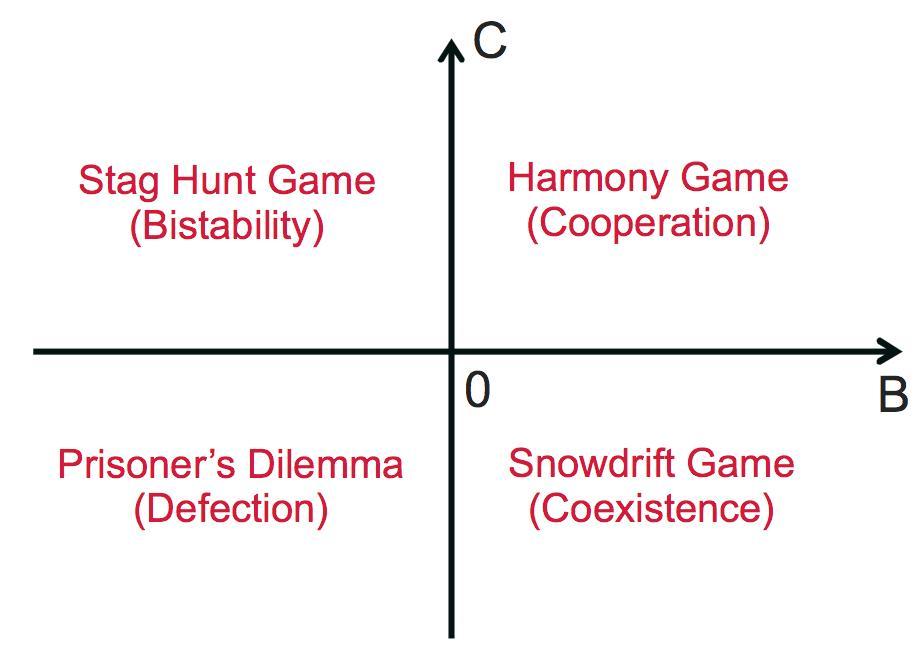}
\end{center}
\caption{ {\bf Illustration of the types and outcomes of a
game-dynamical treatment of \textit{one}-population symmetrical 2x2
games as a function of the two model parameters \textit{B} and
\textit{C }} (see, e.g., \cite{Weibull1995}, pp. 28ff). \textit{B}
and \textit{C }are related to the four payoffs \textit{P},
\textit{R}, \textit{S}, and \textit{T} of symmetrical 2x2 games via
the relations \textit{B = S -- P}  and \textit{C = R -- T}, see Sec.
{\em Methods} for details. The payoff is \textit{R}, if two
interacting individuals show behavior 1, but \textit{P}, if both of
them show behavior 2. When the interaction partners show different
behaviors, the one showing behavior 1 receives the payoff \textit{S}
and the other one the payoff \textit{T}. In the prisoner's dilemmas
(PD), behavior 1 corresponds to cooperation and behavior 2 to
defection. \textit{R} is then called the ``reward'' for mutual
cooperation, \textit{P} the ``punishment'' for mutual defection,
\textit{T} the ``temptation'' for unilateral defection, and
\textit{S} the ``sucker's payoff'' for unilateral cooperation. As
the prisoner's dilemma is characterized by the inequalities
\textit{T} $>$ \textit{R} $>$ \textit{P} $>$ \textit{S}, we have
\textit{B }$<$ 0 and \textit{C} $<$ 0, and defection
(``free-riding'', ``cheating'') is the dominant strategy. For
snowdrift games (SD) with \textit{T} $>$ \textit{R} $>$ \textit{S}
$>$ \textit{P}, which are also called chicken or hawk-dove games, we
have B $>$ 0 and C $<$ 0, and uncooperative behavior is tempting.
The stable stationary solution corresponds to a coexistence of a
fraction $p_0$ = \textbar \textit{B}\textbar /(\textbar
\textit{B}\textbar +\textbar \textit{C}\textbar ) of cooperators
with a fraction $1-p_0$ of defectors (uncooperative individuals).
For harmony games (HG) with \textit{R} $>$ \textit{T }$>$ \textit{S}
$>$ \textit{P}, we have \textit{B} $>$ 0 and \textit{C} $>$ 0, and
everybody will eventually cooperate. Finally, for stag hunt games
(SH) with \textit{R} $>$ \textit{T} $>$ \textit{P} $>$ \textit{S},
which are also called assurance games, we have  \textit{B} $<$ 0 and
\textit{C} $>$ 0. Here, cooperation is uncertain, as the situation
is bistable: If the initial fraction $p_0$ of cooperators is larger
than $p_0$ = \textbar \textit{B}\textbar /(\textbar
\textit{B}\textbar +\textbar \textit{C}\textbar ), everybody is
expected to cooperate in the end, otherwise everybody will
eventually behave uncooperatively. } \label{fig1}
\end{figure}

In order to gain a better understanding of factors preventing or
promoting cooperation among humans or other species, biologists,
economists, social scientists, mathematicians and physicists have
intensively studied game theoretical problems such as the prisoner's
dilemma and the snowdrift game (also known as chicken or hawk-dove
game) \cite{Axelrod1984,Gintis2000,Nowak2006,BenJacob,Griffin}. In
all these games, a certain fraction of people or even everyone is
expected to behave uncooperatively (see Fig. \ref{fig1}). Therefore,
a large amount of research has focused on how cooperation can be
supported by mechanisms such as
\begin{itemize}
\item  repeated interactions \cite{Axelrod1984,HarsanySelten1988,Macy1998,MacyFlache2002},
\item  reputation \cite{RaubWeesie1990,Milinski2002,Castelfranchi1998,Buskens2002},
\item  clusters of cooperative individuals \cite{NowakMay1992,FlacheHegselmann2001},
\item  sanctioning \cite{Heckathorn1990,Kandori1992,BendorMookherjee1990,PosnerRasmussen1999,FehrGachter2002,FehrFischbacher2004b,PLoSBiol},
\item  success-driven migration \cite{HelbingYu2009}, or
\item  economic incentives \cite{Lindbeck1999}.
\end{itemize}

For a discussion and classification of cooperation-promoting mechanisms within an evolutionary game-theoretical framework see Refs.~\cite{Nowak2006, LehmannKeller2006,FletcherDoebeli2009}.

Many game-theoretical studies of social cooperation are based on
models, in which all individuals are assumed to have the same
properties. In reality, however, individuals are different. To
investigate the relevance of this for the resulting outcome and
dynamics of social interactions, we will consider that people of
different gender, status, age, or cultural background may have
heterogeneous preferences (e.g. due to framing effects, see
\cite{Sugden1995,BanarachBernasconi1997,BanarachStahl2000,Banarach2006}.
We will focus here on the case where the preferences are not only
\textit{gradually} different, but where we have two interacting
populations with mutually incompatible preferences, which cannot be
satisfied at the same time. For example, men and women appear to
have incompatible interests many times. Nevertheless, they normally
interact among and between each other on a daily basis. It is also
more and more common that people with different religious beliefs
live and work together, while their religions request some mutually
incompatible behaviors (in terms of the working days and free days,
the food one may eat or should avoid, the headgear, or appropriate
clothing, etc.). A similar situation applies, when people with
different mother tongues meet or businessmen from countries with
different business practices make a deal. In this contribution, we
are interested in identifying factors, which determine whether two
such populations go their own way, find a common agreement, or end
up in conflict \cite{Stouffer1949}. Moreover, we want to understand
the relevance of power in the rivalry of populations
\cite{SaamHarrer1999}.

Our treatment of heterogeneous preferences is based on multi-population games \cite{Cressman1986,Cressman2001,HofbauerSigmund1998,Cressman1995, Cressman1996}. The simpler case where individuals of two \textit{different} populations interact with each other, while they do not interact when belonging to the \textit{same} population, has been nicely summarized by \cite{Weibull1995}, pp. 182ff. An earlier publication on the self-regulation of behavior in animal societies also considers self-interactions within each population \cite{Schuster1981}. In fact, most applications of
multi-population models in evolutionary game theory so far were oriented at the interaction of \textit{biological} species and the study of \textit{ecosystems }\cite{OliveriaFontanari2000,OliveriaFontanari2002,DiederichOpper1989,Sato2005,Galla2006}. Compared to these publications, our treatment focuses on \textit{social} systems, and effects of differences in the power of interacting populations are considered. The problem of conflicts between social groups has been studied before by bimatrix games \cite{EvolGame} such as the ``battle of sexes'' (see, e.g., \cite{HofbauerSigmund1998} and by so-called hypergames (see, e.g., \cite{Kanazawa2007}. However, the related models appear to be less versatile than the one proposed in the following. The main difference to previous approaches is that we study \textit{social} interactions \textit{between} and \textit{within} populations, considering that the power of the involved populations may be different. This generates interesting kinds of system dynamics, which do not appear when self-interactions (between individuals of the same population) are neglected or if all populations are equally strong. For example, we find that it may not only depend on the payoffs, but also on the initial condition, whether the individuals of two populations with incompatible preferences finally show a commonly shared behavior (see Sec. {\em Evolution of normative behavior in the stag hunt game}).

Note that this paper presents more than ``just another model''. First of all, our approach fits particularly well into widely established modeling concepts. Second, it bridges between two different modeling worlds by unifying features of game-theoretical and opinion dynamics models (see Sec. {\em Discussion of previous literature on norms}). Third, the model is analytically tractable \cite{HelbingJohansson2010}. Fourth, it contains very few parameters, while it describes a variety of different system behaviors (although it was not explicitly constructed for this). Despite 3 parameters only (which can be further reduced, since only the signs and quotient of the parameters \textit{B} and \textit{C} matter), the model shows a surprisingly rich behavior and can reproduce a variety of phenomena observed in social systems: (1) the breakdown of cooperation, (2) the coexistence of different behaviors (the establishment of ``subcultures''), (3) the evolution of commonly shared behaviors (``social norms''), and (4) the occurrence of social polarization, conflict, or revolutions. The approach can also be cast into an agent-based model. In fact, agent-based models for the establishment of norms have found a lot of interest, recently \cite{Axelrod1986,ConteCastelfranchi1995,Conte1999,Dignum1999,Epstein2001,Flentge2001,ThebaudLocatelli2001,NakamaruLevin2004,EhrlichLevin2005,Centola2005,GalanIzquierdo2005,Chalub2006, Fent2007,Neumann2008}.

\subsection*{Modeling Approach}

The crucial point of our modeling approach is to adapt the game-dynamical replicator equations for multiple populations \cite{Schuster1981,HofbauerSigmund1998,Helbing1992,HelbingJohansson2010} in a way that reflects interactions between individuals with incompatible preferences (see Sec. {\em Methods}). The resulting equations (\ref{eq1}) and (\ref{eq2}) describe the time evolution of the proportions \textit{p}(\textit{t}) and \textit{q}(\textit{t}) of cooperative individuals in populations 1 and 2, respectively, as individuals imitate more successful behaviors in their own population. Their success depends on the ``payoffs'' quantifying the results of social interactions, i.e., on the own behavior and the behavior of the interaction partner(s).

In order to reflect incompatible interests of both populations, we
assume that population 1 prefers behavior 1, while population 2
prefers behavior 2. If an interaction partner shows the behavior
preferred by oneself, we call this behavior ``cooperative'',
otherwise uncooperative. Accordingly, behavior 1 is cooperative from
the viewpoint of population 1, but uncooperative from the viewpoint
of population 2 (and vice versa). Furthermore, if an individual of
population 1 interacts with an individual of population 2 and both
display the\textit{ same} behavior, we call this behavior
``coordinated''. Finally, if the great \textit{majority} of
individuals in \textit{both} populations shows a coordinated
behavior (in case of a commonly shared behavior), we speak of
``normative behavior'' or a ``behavioral norm''. To establish a
social norm, one of the populations has to give up its preferred
behavior.

What will be the resulting dynamics and outcome of such interactions? Under what conditions will we find ``normative behavior'' (although neither the relative sizes of both populations nor their incompatible preferences change in our model)? To answer these questions, the payoffs from social interactions in 2$\times$2 games are represented by \textit{T}, \textit{R}, \textit{P}, and \textit{S}, as usual (see Sec. {\em Methods} for details). In the prisoner's dilemma, the meaning of these parameters is ``\underline{T}emptation'' to behave non-cooperatively, ``\underline{R}eward'' for mutual cooperation, ``\underline{P}unishment'' for mutual non-cooperative behavior and ``\underline{S}ucker's payoff'' for a cooperative individual meeting an uncooperative one (see Fig. \ref{fig1}). The related game-dynamical replicator equations contain two payoff-dependent parameters, \textit{B = S -- P} and \textit{C = R -- T}.  \textit{C }may be interpreted as gain by coordinating on one's own preferred behavior (if greater than zero, otherwise as loss). \textit{B }reflects the gain when giving up coordinated, but non-preferred behavior. Equations (\ref{eq1}) and (\ref{eq2}) contain a further parameter \textit{f}, which can be interpreted as ``(relative) power'' of population 1 (while 1-\textit{f} would correspond to the relative power of population 2). The relative power may represent the relative size of the populations, but also differences in their material resources (money, weapons, etc.), social capital (status, social influence, etc.), and other factors (charisma, moral persuasion, etc.). It reflects how much influence a population has on the behavioral choice of individuals. When a population has a greater relative power than another one, we call it ``stronger'', if it has less power, we call it ``weaker''. Details of the model and some generalizations are provided in {\em Methods}.

\section*{Results}

We have solved Eqs.~(\ref{eq1}) and (\ref{eq2}) by numerical simulation for different parameter values \textit{B}, \textit{C}, and \textit{f} and different initial conditions \textit{p}(0) and \textit{q}(0). In contrast to the computer-based analysis presented here, a mathematical analysis of the stationary (i.e. time-invariant) solutions and their stability properties has been carried out in a complementary paper \cite{HelbingJohansson2010}. While the linear stability analysis reveals the sensitivity to stochastic fluctuations (random effects, ``noise''), the sensitivity to parameter variations is captured by so-called phase diagrams. Here, we will not go into these technicalities, but rather discuss representative examples of the different \textit{kinds} of system dynamics and their relevance for social systems.

We find that social interactions with incompatible interests do not necessarily produce conflict---they may even promote mutual coordination. Depending on the signs of \textit{B} and \textit{C}, which determine the character of the game, we have four archetypical situations:

\begin{enumerate}
\item  In games like the\textit{ multi-population prisoner's dilemma (MPD), we have B} $<$ 0 and \textit{C} $<$ 0.
\item  In the \textit{multi-population harmony game (MHG)}, we have \textit{B} $>$ 0 and \textit{C} $>$ 0.
\item  \textit{B} $<$ 0 and \textit{C} $>$ 0 applies to games like the \textit{multi-population stag hunt game (MSH)}.
\item  The \textit{multi-population snowdrift game (MSD)} is characterized by \textit{B} $>$ 0 and \textit{C} $<$ 0.
\end{enumerate}

\begin{figure}[!ht]
\begin{center}
\includegraphics[width=4in]{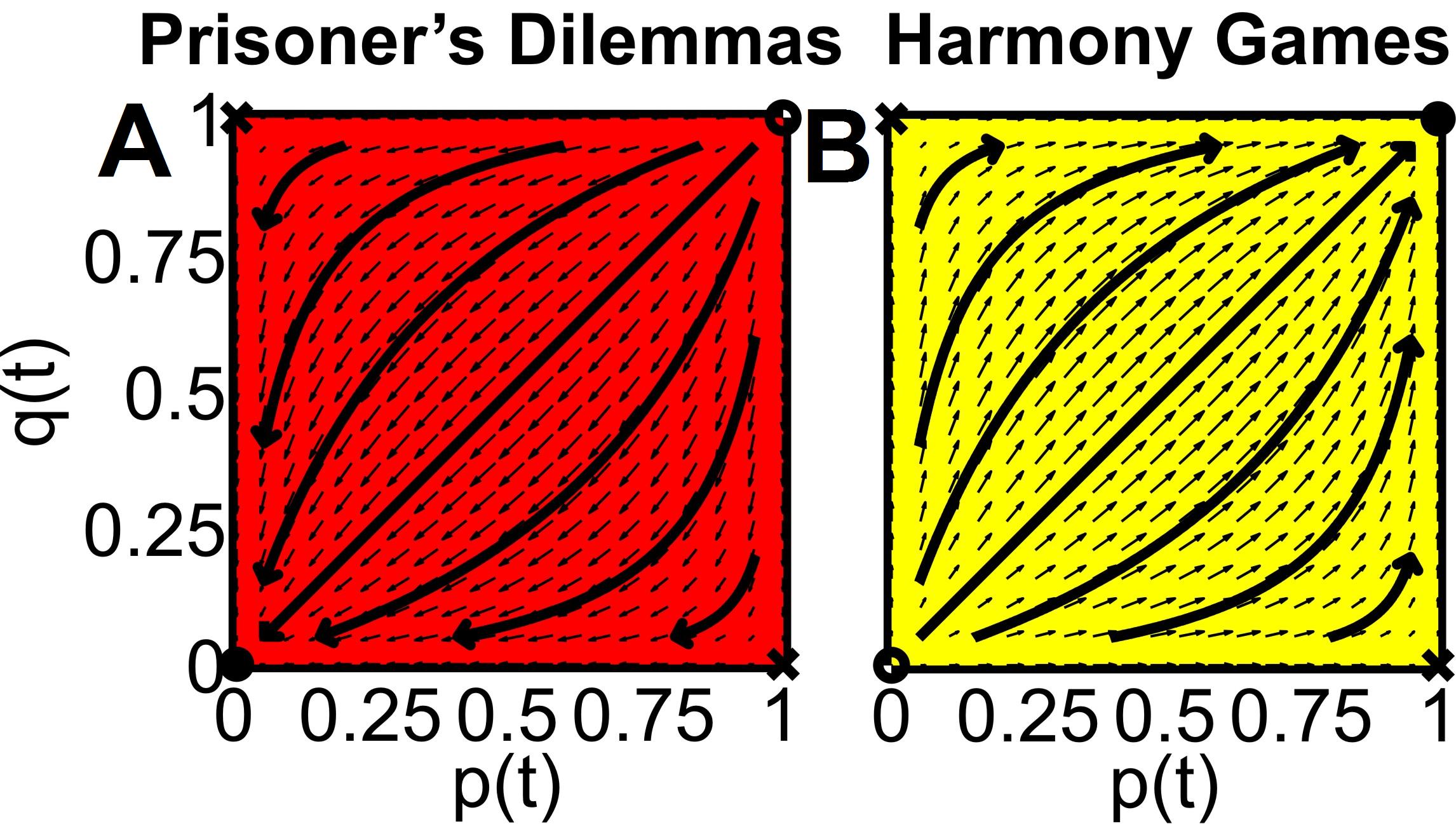}
\end{center}
\caption{ {\bf Simulation results for two interacting populations
with self-interactions and incompatible preferences playing
prisoner's dilemmas or harmony games.} The outcome is visualized by
vector fields (small arrows) and sample trajectories (large arrows)
for \textit{f} = 0.8 (i.e. population 1 is assumed to be stronger
than population 2). \textit{p }denotes the fraction of individuals
in population 1 showing their preferred, cooperative behavior 1, and
\textit{q }is the fraction of cooperative individuals in population
2 showing their preferred behavior 2. A fraction 1 -- \textit{q} of
individuals in population 2 shows the non-preferred behavior 1, and
a fraction 1 -- \textit{p} of individuals in population 1 shows
behavior 2. The vector fields show (d\textit{p}/dt,d\textit{q}/dt),
i.e. the direction and size of the expected temporal change of the
behavioral distribution, if the fractions of cooperative individuals
in populations 1 and 2 are \textit{p}(\textit{t}) and
\textit{q}(\textit{t}). Sample trajectories illustrate some
representative flow lines
(\textit{p}(\textit{t}),\textit{q}(\textit{t})) as time \textit{t}
passes. The flow lines move away from unstable stationary points
(empty circles) and are attracted towards stable stationary points
(black circles). Saddle points (crosses) are attractive in one
direction, but repulsive in another. The colored areas represent the
''basins of attraction'', i.e. all initial conditions
(\textit{p}(0),\textit{q}(0)) leading to the same stable fix point
[red = (0,0), yellow = (1,1)]. Intuitively, the initial conditions
quantify the influence of the previous history. (A) If \textit{B} =
\textit{C} = --1, the individuals in each population are facing
prisoner's dilemma interactions and end up with non-cooperative
behavior. (B) If \textit{B} = \textit{C }= 1, individuals in each
population are playing a harmony game instead, and everybody
eventually behaves cooperatively. The results look similar when the
same two-population games are played with different values of
\textit{f}, \textbar \textit{B}\textbar  or \textbar
\textit{C}\textbar . } \label{fig2}
\end{figure}

In a multi-population prisoner's dilemma with incompatible preferences (MPD), everybody behaves non-cooperatively in the end (see Fig. \ref{fig2}). This does not even change, if one population is stronger than the other one (i.e. $f \ne 1/2$, or if the interaction rate \textit{between} populations is different from the interaction rate \textit{within} populations. This disappointing outcome results despite the fact that non-cooperative behavior in one population corresponds to cooperative one from the perspective of the other. However, as non-cooperative individuals earn a high payoff (the temptation \textit{T}), when meeting a non-cooperative individual of the other population (it is cooperative from the own perspective), there is no incentive to give up defection in their own population---on the contrary.

In contrast, in the multi-population harmony game, everybody finally shows a cooperative behavior in the \textit{own} population, but due to the different preferences, the behaviors in both populations are not coordinated. (Every population just does what it likes, as if both populations had their own ``subcultures'', see Fig. \ref{fig2}).

As we will show in the following, the dynamics and outcome of the
multi- population stag hunt and snowdrift games with incompatible
preferences are more complicated and in marked contrast to the
corresponding one-population games. This can be demonstrated by
systematically exploring the parameter space with computer
simulations. In the following, we will illustrate typical simulation
results by figures and movies showing the stationary solutions (fix
points, evolutionary equilibria) of the games, their basins of
attraction, and representative flow lines. Details are discussed
below and in the captions of Figs. \ref{fig2}--\ref{fig4} (see also
Movies S1--S3 and {\em Methods}).

\subsection*{Evolution of normative behavior in the stag hunt game}

The \textit{one-population} stag hunt game is characterized by an
equilibrium selection problem \cite{Samuelson1998}:
\textit{Everyone} is finally expected to cooperate, if the initial
fraction of cooperative individuals is above $p_0$ = \textbar
\textit{B}\textbar /(\textbar \textit{B}\textbar +\textbar
\textit{C}\textbar ), otherwise \textit{nobody} will behave
cooperatively in the end (see Fig. \ref{fig1}). The same applies to
\textit{non-interacting} populations (see Movie S1). For
\textit{interacting} populations without self-interactions, however,
it \textit{never} happens in the multi-population stag-hunt game
with incompatible preferences that everybody or nobody cooperates in
both populations (otherwise there should be yellow or red areas in
the second part of Movie S2). Although both populations prefer
\textit{different} behaviors, all individuals end up coordinating
themselves on a commonly shared behavior (corresponding to the blue
and green areas in Movie S2). This can be interpreted as
self-organized evolution of a social norm (see below).

\begin{figure}[!ht]
\begin{center}
\includegraphics[width=4in]{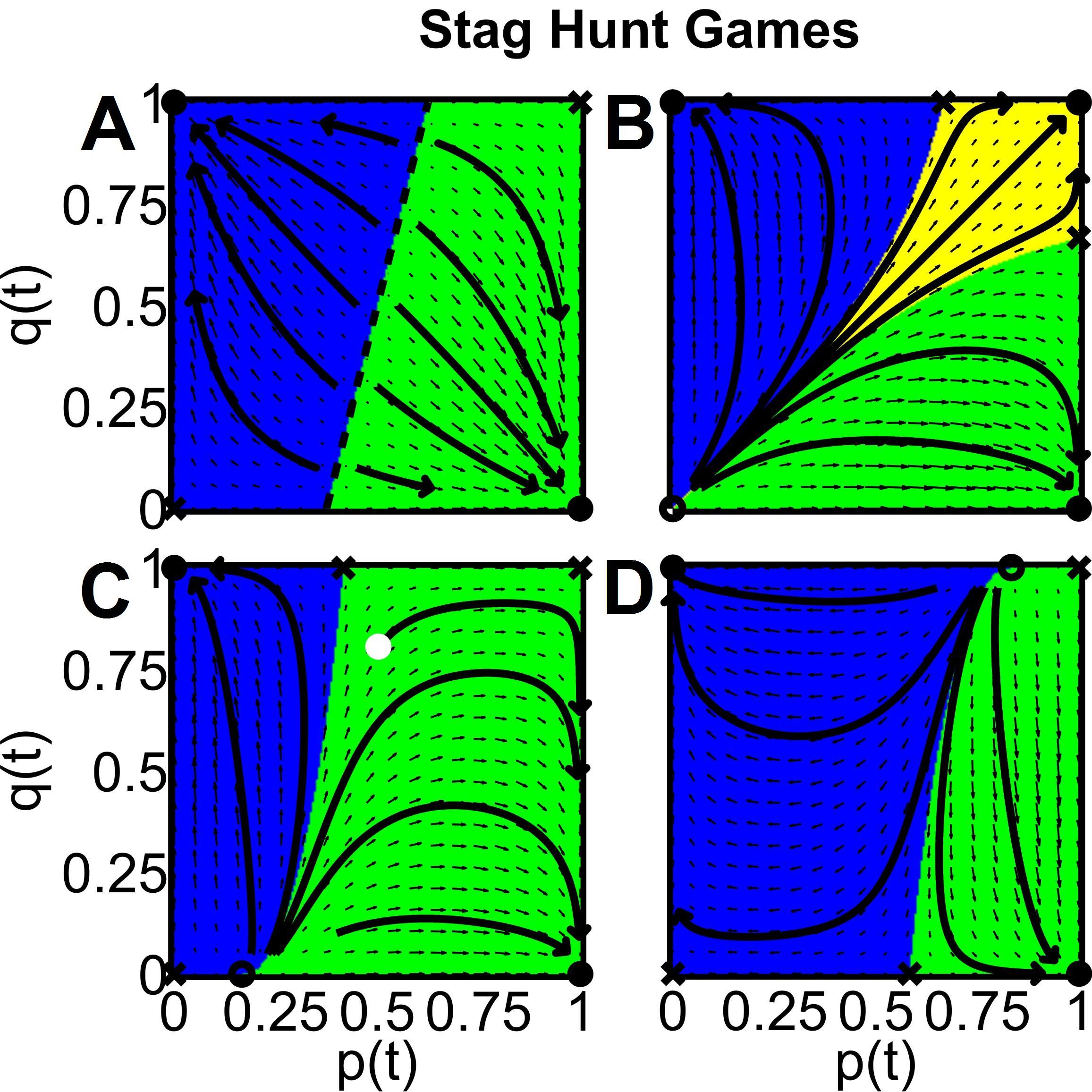}
\end{center}
\caption{ {\bf Simulation results for two interacting populations
with self-interactions and incompatible preferences, playing stag
hunt games.} The corresponding vector fields (small arrows), sample
trajectories (large arrows) and phase diagrams (colored areas) were
determined for \textit{B} $<$ 0 and \textit{C} $>$ 0. The
representation is the same as in Fig. \ref{fig2}. In particular, the
colored areas represent the basins of attraction, i.e. all initial
conditions (\textit{p}(0),\textit{q}(0)) leading to the same stable
fix point (stationary solutions) [yellow = (1,1), blue = (0,1),
green = (1,0)]. The dashed diagonal line represents an infinite
number of unstable fix points. The model parameters are as follows:
(A) \textbar \textit{B}\textbar  = \textbar \textit{C}\textbar  = 1
and \textit{f} = 0.8, i.e. population 1 is more powerful than
population 2, (B) \textbar \textit{C}\textbar  = 2\textbar
\textit{B}\textbar = 2 and \textit{f }= 1/2, i.e. both populations
are equally strong, (C) \textbar \textit{C}\textbar  = 2\textbar
\textit{B}\textbar  = 2 and  \textit{f }= 0.8, (D) 2\textbar
\textit{C}\textbar  = \textbar \textit{B}\textbar  = 2 and
\textit{f} = 0.8. Due to the asymptotically stable fix points at
(1,0) and (0,1), all individuals of both populations finally show
the behavior preferred in population 1, when starting in the green
area, or the behavior preferred in population 2, when starting in
the blue area. This case can be considered to describe the evolution
of a shared behavioral norm. Only for similarly strong populations
(\textit{f} $\approx$ 1/2) and similar initial fractions
\textit{p}(0) and \textit{q}(0) of cooperators in both populations
(yellow area), both populations will end up with population-specific
norms (''subcultures''), corresponding to the asymptotically stable
point at (1,1). The route towards the establishment of a shared norm
may be quite unexpected, as the flow line starting with the white
circle shows: The fraction \textit{q}(t) of individuals in
population 2 who are uncooperative from the viewpoint of population
1 may grow in the beginning, but later on go down dramatically.
Therefore, a momentary trend does not allow one to easily predict
the final outcome of the struggle between two interest groups. }
\label{fig3}
\end{figure}

Note that the previously discussed cases, which neglect interactions
\textit{between} populations or \textit{within} populations, are
applicable to \textit{particular} social systems only. Normally,
however, there are interactions between \textit{different}
populations and, at the same time, interactions between individuals
of the \textit{same} population (``self-interactions''). If this is
taken into account, the case where everybody or nobody cooperates in
both populations\textit{ }is still possible, but it requires that
both populations have similar power ($f \approx ˜1/2$) and that the
initial levels of cooperation, \textit{p}(0) and \textit{q}(0), are
comparable as well. Under such conditions, both populations may
develop separate, coexisting norms (see yellow area in Fig.
\ref{fig3}B and Movie S3), as for the multi-population harmony game.
Normally, however, both populations establish a commonly shared norm
and either end up with behavior 1 (green area in Fig. \ref{fig3}) or
with behavior 2 (blue area).

The  behavior of Eqs.~(\ref{eq1}) and (\ref{eq2}) becomes better
understandable for multi-population games with the payoffs \textit{P
}= \textit{c}, \textit{R }= \textit{b}+\textit{c}, \textit{S }=
\textit{b}, and \textit{T }= 0, where the payoff \textit{b} reflects
the benefit of showing the \textit{preferred} behavior, while
\textit{c} is the payoff for showing \textit{coordinated} behavior
(reflecting the reward for conforming with the behavior of the
interaction partner).  While these payoffs do not correspond to a
stag hunt game, for \textit{c }$>$ 0 and --\textit{c} $<$ \textit{b}
$<$ \textit{c} they \textit{also} lead to \textit{B} =
\textit{b}--\textit{c} $<$ 0 and \textit{C} = \textit{b}+\textit{c}
$>$ 0, which implies exactly the same solutions. One advantage of
this formulation besides the better interpretation is the
possibility to extend it to simultaneous interactions with several
players.
\par
For the sake of illustration of this specification, assume that
individuals of population 1 like to be properly dressed at the beach
and individuals of population 2 enjoy to be naked (\textit{b}$>$0),
but even more than doing what they like, everybody prefers to
conform with the behavior of the interaction partners
(\textit{c}$>$\textit{b}). In situations, when naked people and
those wearing a swimming suit interact with each other at the same
part of the beach, our equations allow one to identify conditions
under which one population eventually sets the behavioral standards
or under which a mixture of both behaviors persists. (In contrast to
the related multi-population game \textit{without}
self-interactions, see Movie S2, there is a possibility that the two
populations do \textit{not }coordinate their behaviors, and
everybody ends up doing what he or she likes.) Note that, if both
populations interact in \textit{space}, dressed people and nudists
may segregate, even when there was no disapproval between both
behaviors. As a consequence, there may be different
``(sub-)cultures'' in different parts of the beach, as is often
observed. This becomes understandable by the circumstance that the
effect of mutual \textit{disapproval} of the non-preferred behavior
(which may be described by the payoffs \textit{P }= \textit{0},
\textit{R }= \textit{b}, \textit{S }= \textit{b--c}, and \textit{T
}= --c) \textit{again} leads to \textit{B} = \textit{b}--\textit{c}
$<$ 0 and \textit{C} = \textit{b}+\textit{c} $>$ 0. Therefore, the
same kind of dynamics results as in the case where there is a
tendency to conform with others.

In conclusion, due to the payoff structure of the multi-population stag hunt game and other multi-population games with \textit{B} $<$ 0 and \textit{C} $>$ 0, it can be profitable to coordinate oneself with the prevailing behavior in the other population. Yet, the establishment of a norm requires the individuals of one population to give up their\textit{ own} preferred behavior in favor of the one preferred by the\textit{ other} population. Therefore, it is striking that the preferred behavior of the\textit{ weaker} population can actually win through and finally establish the norm (see blue areas in Figs. \ref{fig3}A,C,D). \textit{Who} adapts to the preferred strategy of the other population essentially depends on the\textit{ initial} fractions of behaviors (and, thereby, on the previous history). The majority behavior in the beginning is likely to determine the resulting behavioral norm, but a powerful population is in a favorable position: The area of possible histories leading to an establishment of the norm preferred by population 1 tends to increase with power \textit{f} (compare the size of the green areas in Figs. \ref{fig3}B+C).

\subsection*{Discussion of the equilibrium selection problem}

As was indicated already, when two populations with incompatible
preferences interact among and between each other,  the behavior of
the stag hunt game changes completely: Then, the values of the
payoff-dependent parameters \textit{B} and \textit{C} have an
influence on the stable stationary solutions, and inner stationary
points $(p,q)$ with $0<p,q<1$ disappear, if $|B|\ne |C|$
\cite{HelbingJohansson2010}.

It is noteworthy that, {\it without} interactions between populations, the stag hunt game implies an interesting equilibrium selection problem \cite{HarsanySelten1988,Risk,Risk1,Risk2}, since it has {\it several} stable solutions. These are classified as payoff-dominant solution (which maximizes the individual payoff, if the interaction partner decides in the same way) and risk-dominant solution (which ``minimizes the maximum damage'', i.e. maximizes the individual payoff for the worst-case choice of the interaction partner and corresponds to non-cooperative behavior). Which of these solutions is selected depends on the initial conditions, and there is a monotonous increase or decrease of the fraction $p(t)$ of cooperative individuals in the course of time $t$. In the one-population stag hunt game, the payoff-dominant solution corresponds to cooperative behavior by everybody ($\lim_{t\rightarrow \infty} p(t) = 1$). It is selected, if the initial fraction $p(0)$ of cooperative individuals is above the value $p_0 = |B|/(|B|+|C|)$ of the unstable stationary solution. Instead, the risk-dominant solution results for $p(0)<p_0$ and corresponds to non-cooperative behavior by everyone ($\lim_{t\rightarrow \infty} p(t) = 0$).

In the multi-population games with interactions {\it across} populations, the risk-dominance concept is not sufficient to understand the dynamics and outcome of the game. For example, when two populations with incompatible preferences play stag hunt games {\it without} self-interactions, the game is of the ``battle of sexes'' type, and  there are no thresholds that would separate payoff-dominant from risk-dominant solutions \cite{HofbauerSigmund1998,HelbingJohansson2010}. When both, interactions {\it and} self-interactions are considered, the inner stationary point disappears whenever $|B|\ne|C|$. The unstable solution is rather located at the boundary or in one of the corners. Moreover, as Figs. \ref{fig3}C and \ref{fig3}D illustrate, the incompatibility of preferences can destabilize the solutions (0,0) or (1,1). Even an initial increase in the fractions $p(t)$ and $q(t)$ of cooperative individuals (i.e. $p'(0) > 0$ and $q'(0) >0$, where $p'(t) = dp(t)/dt$ and $q'(t)=dq(t)/dt$) does not imply that the system will end up in the stationary solution (1,1). While there still exists a payoff- and a risk-dominant solution for the stronger population, there is no threshold behavior for the weaker population, as Figs. \ref{fig3}B+C show.

\begin{figure}[!ht]
\begin{center}
\includegraphics[width=4in]{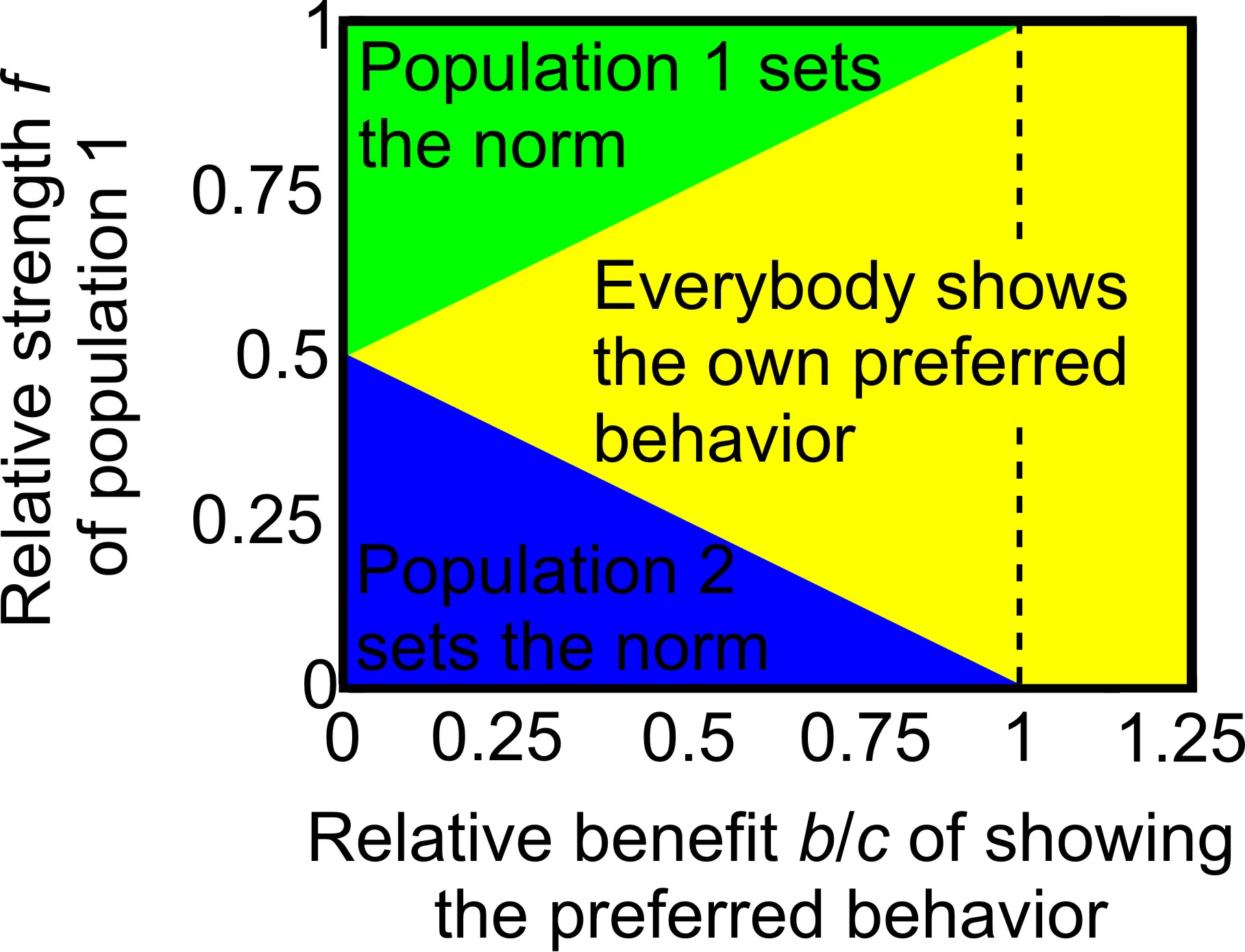}
\end{center}
\caption{ {\bf Illustration of the finally resulting system state
for the two-population stag hunt game with interactions and
self-interactions, and incompatible preferences.} The outcome is
displayed as a function of the relative strength $f$ of population 1
and the ratio $b/c$ between the benefit $b$ of showing the preferred
behavior and the benefit $c$ of showing coordinated behavior. If
$b>c$, individuals always show the behavior they prefer. However, if
$b\le c$, a commonly shared behavior (i.e. a ``social norm'') may
result, depending on the respective initial conditions. The initial
commitments $p(0)$ and $q(0)$ to the respectively preferred behavior
determine also, who sets the norm. If all individuals initially show
the behavior they prefer (i.e. $p(0)=q(0)=1$, as assumed in the
above illustration), the stronger population sets the norm, given
the difference between both population strengths, $f$ and $(1-f)$,
is larger than a certain threshold. Otherwise, individuals in each
population show their preferred behavior. The threshold depends on
the relative strength $f$ and the size of $b/c$, i.e. the fraction
of the relevant payoffs $b$ and $c$. It can, in principle, be
determined from analytical results derived in Ref.
\cite{HelbingJohansson2010}.}\label{new}
\end{figure}

In the limiting case where the relative size $f$ of population 1 goes to one, the resulting fraction of cooperative individuals in population 2 is completely determined by the initial fraction $p(0)$ of cooperative individuals in population 1, while the initial fraction $q(0)$ of cooperative individuals in population 2 does not have any influence. No matter whether population 1 selects the payoff-dominant solution (for large enough values of $p(0)$) or the risk-dominant solution (for small values of $p(0)$), the behavior in population 2 is always coordinated with population 1 in the end \cite{HelbingJohansson2010}. The risk-dominant case can be interpreted such that population 2 effectively manages to set the norm. Figure \ref{new} shows the parameter dependence for the general case, using the alternative parametrization $B=b-c$ and $C=b+c$, where $b$ is the benefit of showing the preferred behavior and $c$ is the benefit of showing coordinated behavior. The classical coordination game, where individuals {\it always} form a behavioral convention (i.e. coordinate on a behavioral norm) results for $b=0$ \cite{Helbing1992}.

\par

\subsubsection*{Discussion of previous literature on norms}

Note that the subject of social norms is a multi-faceted research
field, and there is no single, generally accepted definition of what
norms are \cite{Voss2001,Opp2001,Bicchieri2006,Bicchieri2009}.
Definitions range from the concept of ``oughtness''
\cite{Homans1974} to a behavioral regularity or shared behavior with
a sanctioning (``punishment'') of non-conforming behavior
\cite{Popitz1980,Ellickson1991,CialdiniTrost1998,Ostrom2000,FehrGachter2000,Horne2001a,Horne2008,Boyd2003,FehrFischbacher2004a,FehrFischbacher2004b,Henrich2006}.
However, not all authors agree on the necessity of the sanctioning
element \cite{Elster1989a,Elster1989b}. In our manuscript, we define
``normative behavior'' or a ``behavioral norm'' as a situation in
which behaviors are shared among a large majority of individuals.

The question of how behavioral consensus may evolve has been addressed by opinion dynamics models such as voter models \cite{EhrlichLevin2005} or models of in- and out-group interactions \cite{Kitts2006,Fent2007}. The currently most common approach to behavioral norms is based on game theory and relates the issue to ultimatum games \cite{Elster1989a,Samuelson1998,Bicchieri2006}, stag hunt games \cite{Skyrms1996,Skyrms2003}, prisoner's dilemmas \cite{Ullmann-Margalit1977,Axelrod1986,Heckathorn1990,Ellickson1991,BendorSwistak2001}, or related concepts \cite{Chalub2006}. For overviews see \cite{Ullmann-Margalit1977,Voss2001,HechterOpp2001}. Yet, the majority of these models investigate conditions under which people comply with a\textit{ preset} norm. Repeated interactions \cite{Axelrod1984} and the sanctioning of non-conforming behavior \cite{Fehr2002} are two such conditions. It was less clear, however, whether and how a commonly shared norm would be established in situations, where the involved populations prefer \textit{different} norms.

Our own approach to understanding behavioral consensus relates to
evolutionary game theory, but in contrast to most models for the
evolution of norms, it assumes a heterogeneity of individual
preferences and therefore involves several populations. In {\em
Methods} and {\em Discussion}, we furthermore consider the role of
sanctioning as a mechanism that supports the evolution of norms, but
point out that there are other mechanisms which are expected to
support the evolution of norms as well. Note that the term
``evolution'' is used by us in the sense of ``temporal evolution''
or ``eventual establishment'', not in the sense of ``biological
evolution'' or ``spontaneous emergence''.

\subsubsection*{Examples and classification of norms}

In the following, we will illustrate the concept of social norms by some examples. In his book ``The Cement of Society'', Jon Elster \cite{Elster1989b} discusses consumption norms, norms against behavior `contrary to nature', norms regulating the use of money, norms of reciprocity, medical ethics, codes of honor, norms of retribution, work norms, norms of cooperation, and norms of distribution. Norms underlying common neighborhood or business practices have been analyzed by Macaulay \cite{Macaulay1963}, Ostrom \cite{Ostrom1990} and Ellickson \cite{Ellickson1991}.

When discussing norms, it is useful to distinguish between coordination norms and cooperation norms \cite{Ullmann-Margalit1977}. \textit{Coordination norms} are self-enforcing. They are established, when it is advantageous for people to show a coordinated behavior, but it does not matter which of the behavioral options people agree upon. In that case, one also speaks of behavioral conventions, and one-population models are often sufficient to describe the underlying dynamics \cite{Helbing1992,Young1993}. Examples of conventions are the preference of pedestrians to walk on one side \cite{Helbing1991,Helbing1992,Helbing1994,Helbing1996,Moussaid2009} (for example, the right-hand side in continental Europe or the left-hand side in Japan), the direction of writing, the way people greet each other (whether one gives a hand and which one, whether one hugs or kisses the person and how many times), the way people eat, the color of clothes worn by political movements, and signs used by followers of certain ideas or tastes to identify each other (e.g. tattoos or hanky codes).

In contrast to coordination norms, \textit{cooperation norms} are not self-enforcing, since there are incentives for unilateral deviance. In our paper, we analyze situations, where people have \textit{ different preferences}, so that at least a certain fraction of people is tempted to show non-conforming behavior. Gender norms may serve for illustration. Just imagine a ``battle of the sexes'' in a \textit{group} of friends (rather than between two players), where men prefer to watch soccer and women prefer to see a cultural performance, to discuss a stereotypical example. Note that, in our model, interactions occur not only between men and women, but also among men and among women, so the outcome will depend on their relative power.

Religious norms constitute another case, where people with incompatible preferences interact with each other. A similar thing applies to legal norms, when people believing in a pluralistic civil law system interact with people believing in a religious law system. It is well-known that these law systems have incompatible implications with regard to certain issues. A similar situation applies, when businessmen from countries with different business practices make a deal or people with different mother languages meet. In our opinion, communicating in a language is not just a coordination problem. Most people have a clear preference for their mother tongue, and it shapes even the way of thinking and of social interactions. Therefore, when people with different mother tongues meet, there is an incentive to unilaterally deviate from speaking the same language (e.g., due to differences in proficiency). Nevertheless, a common (``normative'') language \textit{can} establish, as is impressively shown, for example, by the unification of regionally spoken dialects in Germany triggered by the Luther bible. Note, however, that proper language use does not seem to be fully self-enforcing, otherwise lexica, schools, and related legal regulations would not be needed.

Besides coordination and cooperation norms, it appears to make sense to distinguish a third class of ``\textit{hybrid norms''}, which share features of both kinds of norms. This case occurs when it is \textit{costly }to switch the behavior (i.e. when transaction costs are high). Technological norms may serve as an example. Customers will usually profit from shared technical standards concerning, for example, the type of keyboard (QWERTY or Dvorak) \cite{David1985}, the kind of operating system (Windows vs. Mac OS or Linux), the technology of video players (VHS vs. Beta MAX) \cite{Arthur1989} or high resolution DVD players (blue-ray vs. HD DVD). In such cases, customers do not have incentives to deviate from a technological standard, once it has established everywhere. In the beginning, however, a common standard does not evolve by itself, as customers buy different technologies and are reluctant to give up the technology they have invested in. Therefore, the use of a single technology is not self-enforcing in the beginning. Once there is a majority standard, however, most people will join it after some time, and their preferences change accordingly.

One should also mention that \textit{some} conventions or norms are set by law, e.g. the driving side \cite{Young1993}, or may involve an intentional segregation from other groups (e.g. when groups develop their own dress-codes or symbols such as tattoos). This touches issues of group dynamics, which are beyond the scope of this paper. The novel contribution of our model is that it sheds new light on the problem of whether a norm can establish (under what conditions) and how (in terms of the dynamics). There are even exact mathematical results for this \cite{HelbingJohansson2010}. In particular, our model reveals that the dynamics and finally resulting state of the system is not only determined by the payoff structure. It also depends on the power of populations and even on the initial proportions of cooperative individuals (the initial conditions or previous history).

Within our model of the evolution of norms, one could say that Figs. \ref{fig3}A,C,D represent the formation of \textit{coordination norms}, as one behavioral norm is \textit{always} established (reflecting self-enforcement). Figure \ref{fig3}B, in contrast, describes situations where two different behaviors can coexist in a stable way (see the yellow basin of attraction). Due to this lack of self-enforcement, it makes sense to attribute this case to the problem of establishing a \textit{cooperation norm}. This relevant case can \textit{only} occur, when taking into account self-interactions in multi-population games. It is also interesting to note, that the application of group pressure can transform the problem of establishing a \textit{cooperation} norm into the problem of establishing a \textit{coordination} norm (see Sec. {\em Methods}). Finally, the case of \textit{hybrid norms} can be treated by considering transaction costs in our model.

\subsection*{Occurence of social polarization in the snowdrift game}

Let us now turn our attention to the discussion of snowdrift games. In the \textit{one-population} case, there is \textit{one} stable stationary point, corresponding to a fraction $p_0$ = \textbar \textit{B}\textbar /(\textbar \textit{B}\textbar +\textbar \textit{C}\textbar ) of cooperative individuals (see Fig. \ref{fig1}). If this would be transferable to the multi-population case we are interested in, we should have \textit{p} = \textit{q} = $p_0$ in the limit of large times $t \rightarrow \infty$. Instead, we find a variety of different outcomes, depending on the values of the model parameters \textit{B}, \textit{C}, and \textit{f }(see Fig. \ref{fig4}):

\begin{figure}[!ht]
\begin{center}
\includegraphics[width=4in]{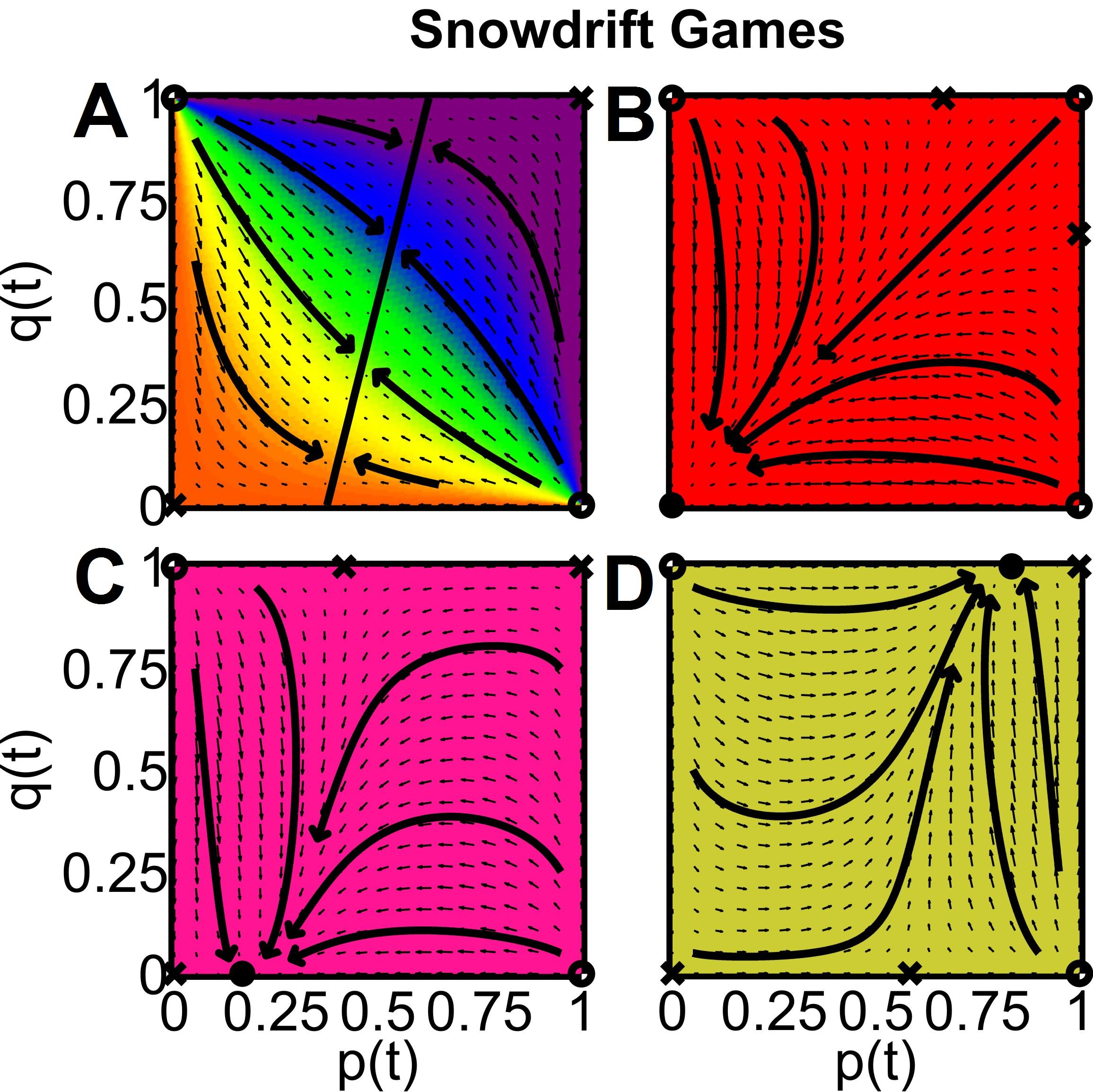}
\end{center}
\caption{{\bf Two interacting populations with self-interactions and
incompatible preferences, playing snowdrift games.} The
corresponding vector fields (small arrows), sample trajectories
(large arrows) and phase diagrams (colored areas) were determined
for \textit{B} $>$ 0 and \textit{C} $<$ 0. The flow lines move away
from unstable stationary points (empty circles) and are attracted
towards stable stationary points (black circles and solid diagonal
line). Saddle points (crosses) are attractive in one direction, but
repulsive in another. The representation is the same as in Fig.
\ref{fig2}. In particular, the colored areas represent the basins of
attraction, i.e. all initial conditions
(\textit{p}(0),\textit{q}(0)) leading to the same stable fix point
[red = (0,0), salmon = (\textit{u},0), mustard = (\textit{v},1),
rainbow colors = (\textit{u},\textit{v}), with 0 $<$
\textit{u},\textit{ v} $<$ 1]. The model parameters are as follows:
(A) \textbar \textit{B}\textbar  = \textbar \textit{C}\textbar  = 1
and \textit{f} = 0.8, i.e. population 1 is more powerful than
population 2, (B) \textbar \textit{C}\textbar  = 2\textbar
\textit{B}\textbar  = 2 and \textit{f} = 1/2, i.e. both populations
are equally strong, (C) \textbar \textit{C}\textbar  = 2\textbar
\textit{B}\textbar  = 2 and \textit{f} = 0.8, (D) 2\textbar
\textit{C}\textbar  = \textbar \textit{B}\textbar  = 2 and
\textit{f} = 0.8. (A) In the multi-population snowdrift game (MSD),
a mixture of cooperative and uncooperative behaviors results in both
populations, if \textbar \textit{B}\textbar =\textbar
\textit{C}\textbar . (B) For \textbar
\textit{B}\textbar$<$\textbar\textit{C}\textbar  and equally strong
populations, everybody ends up with non-cooperative behavior in each
population. (C) For \textbar \textit{B}\textbar  $<$ \textbar
\textit{C}\textbar  and $f - 1/2\gg 0$, the weaker population 2
forms a ``tacit alliance'' with the minority of the stronger
population 1 and opposes its majority. (D) Same as (C), but now, all
individuals in the weaker population 2 show their own preferred
behavior after the occurrence of a ``revolutionary'' transition,
during which the stable stationary solution (the evolutionary
equilibrium) changes discontinuously from (\textit{u},0) to
(\textit{v},1). } \label{fig4}
\end{figure}

\begin{enumerate}
    \item[(a)]  \textit{The interactions between both populations shift the fraction of cooperative individuals in each population to values different from $p_0$.} If \textbar \textit{B}\textbar  = \textbar \textit{C}\textbar , we discover a line of \textit{ infinitely many} stationary points, and the actually resulting stationary solution uniquely depends on the initial condition (see Fig. \ref{fig4}A). This line satisfies the relation \textit{q} = \textit{p} only if \textit{f} = 1/2, while for most parameter combinations we have \textit{q} $\neq$ \textit{p} $\neq$ $p_0$. Nevertheless, the typical outcome in the case \textbar \textit{B}\textbar  = \textbar \textit{C}\textbar  \ is characterized by a finite fraction of cooperative individuals in each population.
    \item[(b)]  \textit{Conflicting interactions between two equally strong groups destabilize the stationary solution q = p = $p_0$ of the one- population case, and both populations lose control over the final outcome.} For \textbar \textit{B}\textbar  $\neq$ \textbar \textit{C}\textbar , all stationary points are \textit{ discrete} and located on the \textit{ boundaries}, and only\textit{ one} of these points is an evolutionary equilibrium. If both populations have\textit{ equal} power (\textit{f} = 1/2), we either end up with non-cooperative behavior by everybody (if $p_0<1/2$, see Fig. \ref{fig4}B), or everybody is cooperative (if $p_0>1/2$). Remarkably, there is no mixed stable solution between these two extremes.
    \item[(c)]  \textit{The stronger population gains control over the weaker one, but shows polarization itself, and a change of the model parameters may induce a revolution.} If \textbar \textit{B}\textbar  $\neq$ \textbar \textit{C}\textbar \ and population 1 is much\textit{ stronger} than population 2 (i.e., $f - 1/2 \gg 0$), we find a finite fraction of cooperative individuals in the stronger population, while either 0\% or 100\% of the individuals are cooperative in the weaker population. A closer analysis reveals that the resulting overall fraction of cooperative individuals fits exactly the expectation $p_0$ of the stronger population \cite{HelbingJohansson2010}, while from the perspective of the weaker population, the overall fraction of cooperative individuals is largely different from $p_0$= \textbar \textit{B}\textbar /(\textbar \textit{B}\textbar +\textbar \textit{C}\textbar ). Note that the stronger population alone can not reach an overall level of cooperation of $p_0$. The desired outcome can only be produced by effectively \textit{controlling} the behavior of the weaker population. This takes place in an unexpected way, namely by \textit{polarization}: The stronger population splits up into fractions of people showing different behaviors, which may give rise to social differentiation, inequality, and conflict. In the weaker population 2, \textit{ }everyone either shows behavior 1 (namely for $p_0<1/2$, see Fig. \ref{fig4}C), otherwise everyone shows behavior 2 (see Fig. \ref{fig4}D). There is no solution in between these two extremes (apart from the special case $p_0=1/2$ for \textbar \textit{B}\textbar  = \textbar \textit{C}\textbar ).
\end{enumerate}

It comes as a further surprise that the behavior in the weaker population is always coordinated with the \textit{minority} behavior in the stronger population. Due to the payoff structure of the multi- population snowdrift game, it is profitable for the weaker population to oppose the majority of the stronger population, which creates a tacit alliance with its minority. Such antagonistic behavior is well-known from protest movements \cite{Opp2009} and finds here a natural explanation.

Moreover, when \textbar \textit{C}\textbar \ changes from values
greater than \textbar \textit{B}\textbar \ to values smaller than\
\textbar \textit{B}\textbar , there is an unexpected, discontinuous
transition in the weaker population 2 from a state in which
everybody is cooperative from the point of view of population 1 to a
state in which everybody shows the\textit{ own} preferred behavior 2
(see Movie S3). History and science \cite{Kuhn1962} have seen many
abrupt regime shifts of this kind. Revolutions caused by class
conflict provide ample empirical evidence for their existence.
Combining the theory of phase transitions with ``catastrophe
theory'' \cite{Zeeman1977} offers a quantitative scientific approach
to interpret such revolutions as the outcome of social interactions
\cite{WeidlichHuebner2008}. Here, their recurrence becomes
understandable in a unified and simple game-theoretical framework.

\section*{Discussion}

Multi-population game-dynamical replicator equations provide an
elegant and powerful approach to study the dynamics and outcomes
expected for populations with incompatible interests. A detailed
analysis reveals how combining interactions within and between
populations and considering differences in their power can
substantially change the dynamics of various game theoretical
dilemmas (compare Movies S2 and S3 with Movie S1). Generalizations
to \textit{more} than 2 behaviors or groups and to different payoffs
for in- and out-group interactions are easily possible (see Sec.
{\em Methods}).

When two populations with incompatible preferences interact among
and between each other, we find the same stationary points for the
prisoner's dilemma and the harmony game as for the corresponding
non-interactive games. In particular, interactions across
populations do not change the attractive solutions (Nash equilibria)
of these games. However, the behavior of the snowdrift game and the
stag hunt game changes completely, and their dynamics is
particularly interesting. For the interactive case, the signs of the
payoff-dependent parameters \textit{B} and \textit{C} do not only
determine the character of the game, but also the location and
stability of the stationary solutions, and the basins of attraction.

In the multi-population snowdrift game, for example, there is a
discontinuous (``revolutionary'') transition, when 1-\textbar
\textit{B}\textbar /\textbar \textit{C}\textbar  changes its sign.
On top of this, the power \textit{f} has a major influence on the
outcome, and the initial distribution of behaviors (and,
consequently, the previous history) can be crucial, also for the
multi-population stag hunt game. Note that such a rich system
behavior is already found for the \textit{simplest} setting of our
model and that the concept of multi-population game-dynamical
equations may be generalized in various ways to address a number of
challenging questions in the future: How can we gain a better
understanding of a clash of cultures, the outbreak of civil wars, or
conflicts with ethnic or religious minorities? How can we
analytically study migration and group competition? When do social
systems become unstable and experience a polarization of society?
How can we understand the emergence of fairness norms in bargaining
situations?

Another interesting aspect of our model is that it makes a variety
of quantitative predictions.  Therefore, it could be tested
experimentally with iterated games in the laboratory, involving
several groups of people with random matching and sufficiently many
iterations. Suitable changes in the payoff matrices should be able
to confirm the mathematical conditions under which different
archetypical types of social phenomena or discontinuous transitions
in the system behavior can occur: (1) the breakdown of cooperation,
(2) in-group cooperation (the formation of ``sub-cultures''), (3)
the evolution of shared behavioral norms, and (4) societal
polarization or conflict with the possibility of a revolutionary
regime shift. The findings are particularly important to understand
interactions between human populations with different ethnic,
cultural or religious backgrounds. However, they are also relevant
for social features within animal societies
\cite{Schuster1981,Brock,Wilson,Bonner} or even for interactions
among bacteria \cite{BenJacob,Griffin}.

\begin{figure}[!ht]
\begin{center}
\includegraphics[width=4in]{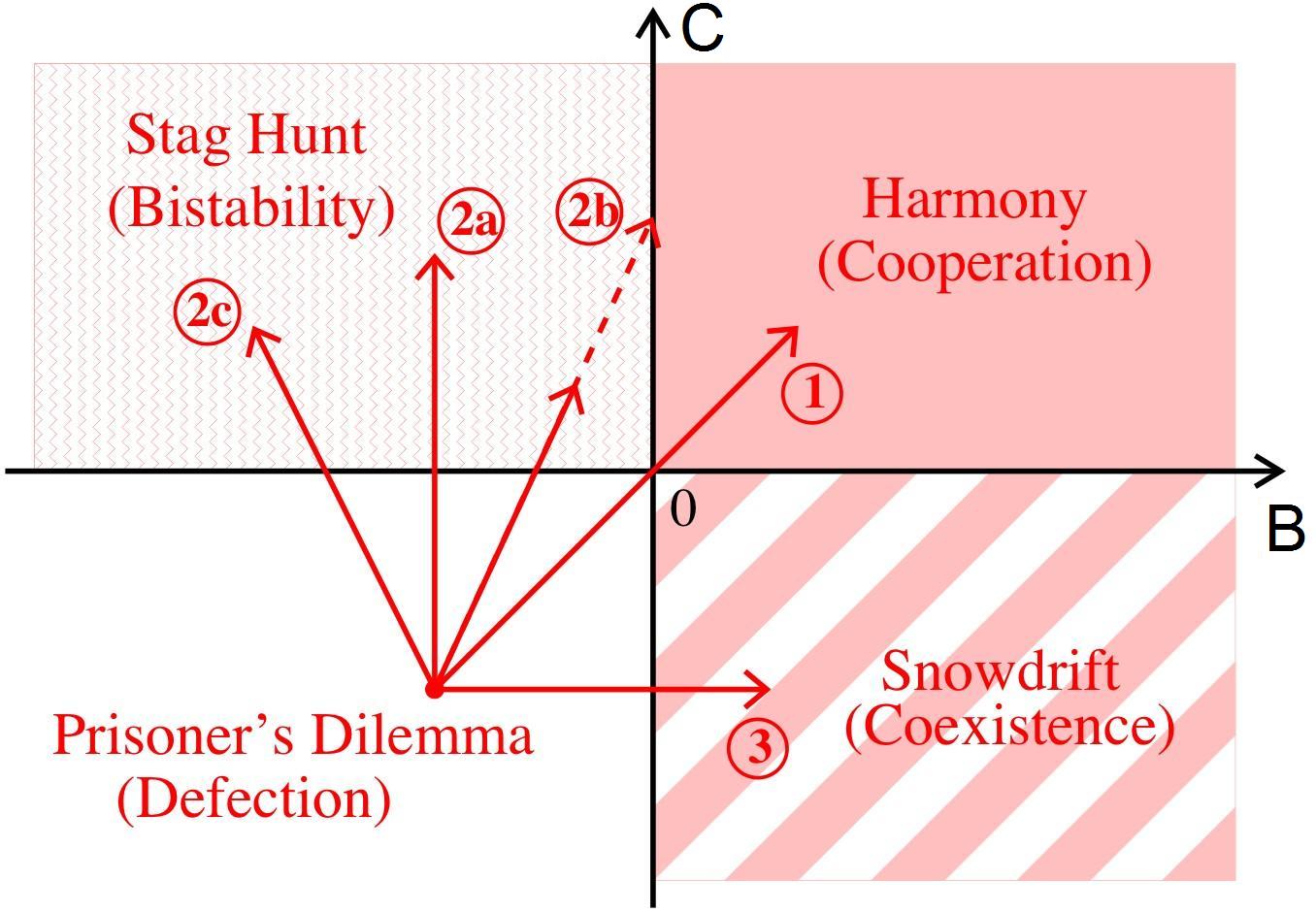}
\end{center}
\caption{ {\bf Illustration of different routes to cooperation
(through arrows), assuming particular reproduction-selection
mechanisms \cite{Nowak2006}.} The direction of the arrows can be
mathematically calculated \cite{HelbingLozano2010}. Route 1 reflects
the way in which the payoff-dependent parameters \textit{B} and
\textit{C} of the game (see Fig. \ref{fig1}) are effectively
modified by kin selection (genetic relationship), network
reciprocity (clustering of individuals showing the same behavior),
or group selection (competition between different groups). Route 2a
corresponds to the effect of direct reciprocity (due to the ``shadow
of the future'' through the likelihood of future interactions).
Route 2b belongs to the mechanism of indirect reciprocity (based on
reputation effects), and route 2c reflects costly punishment. Route
3 results for certain kinds of network interactions
\cite{Ohtsuki,Nowak2006}. } \label{fig5}
\end{figure}

The significant influence of the respective payoffs of social
interactions on the resulting outcome has crucial implications for
society, law and economics
\cite{Sugden1986,Sugden1989,Sugden1998,KofordMiller1991,Sethi1996,Binmore1994,Binmore2005,Samuelson1998,Platteau2000,Ellickson2001,Bohnet2001}.
There, conflicts need to be avoided or solved, and norms and
standards are of central importance. For society, norms are equally
important as cooperation, since they reduce uncertainty, bargaining
efforts, and (potentially) also conflict in social interactions.
They are like social forces guiding our interactions in numerous
situations and subtle ways, creating an ``invisible hand'' kind of
self-organization of society \cite{Axelrod1986}. Nevertheless, their
ubiquity is quite surprising, as norms require people to constrain
self-interested behavior \cite{Keizer2008} and to perform socially
prescribed roles. Yet, widespread cooperation-enhancing mechanisms
such as direct reciprocity due to repeated interactions
\cite{Axelrod1984} and indirect reciprocity based on reputation
\cite{Milinski2002} can transform a prisoner's dilemma into stag
hunt interactions, see Fig. \ref{fig5}
\cite{Skyrms2003,Nowak2006,HelbingLozano2010}.

\begin{figure}[!ht]
\begin{center}
\includegraphics[width=4in]{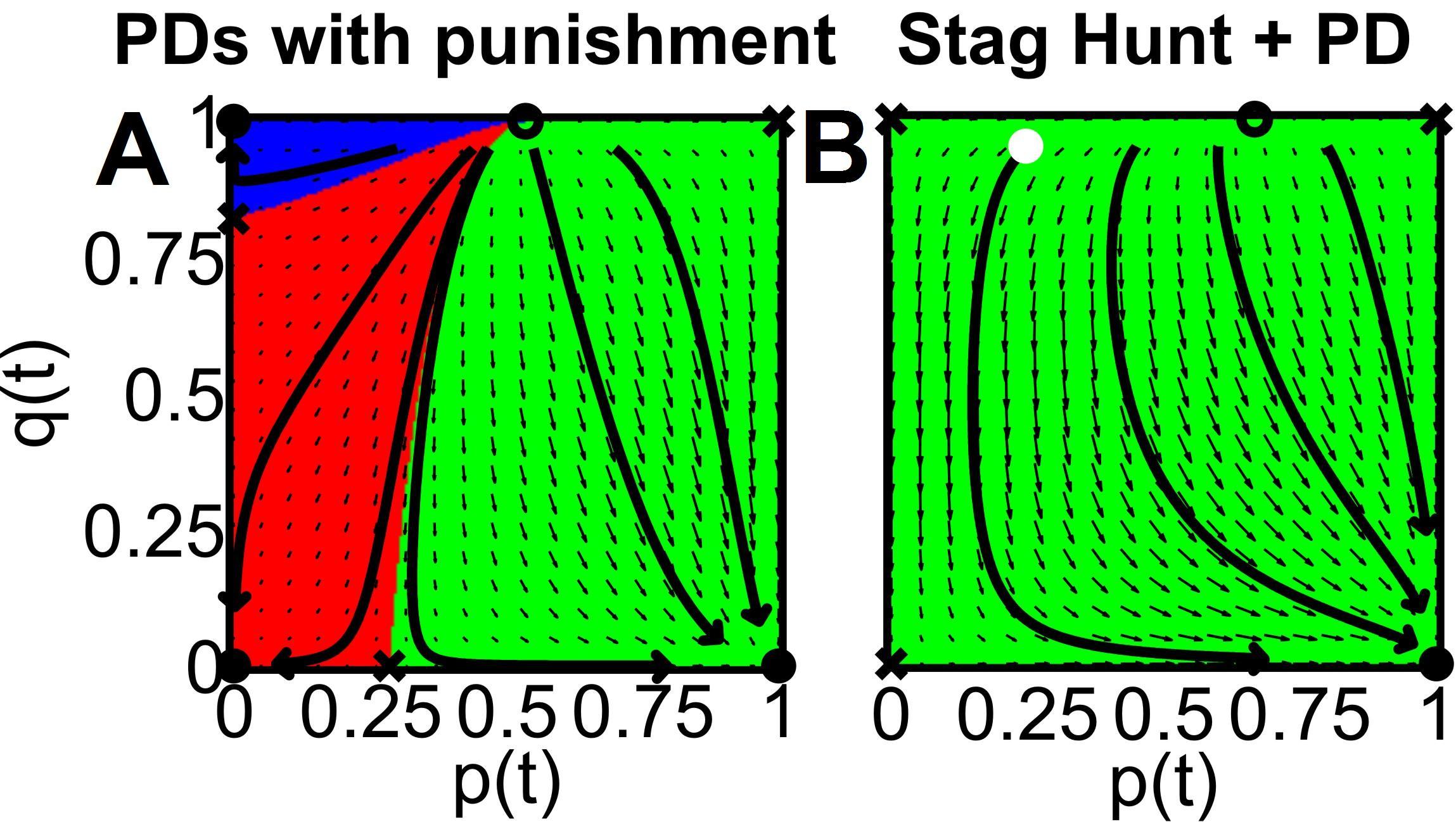}
\end{center}
\caption{ {\bf Illustration of different ways to establish a
behavioral norm.} (A) Prisoner's dilemma games with model parameters
\textit{B} = -1, \textit{C} = -2, when population 1 is much stronger
than population 2 (\textit{f }= 0.8) and individuals of both
populations apply costly punishment, whenever the respective focal
individual behaves cooperatively, but (from his or her perspective)
the interaction partners does not \cite{Traulsen2009}. The
punishment cost was assumed to be $\beta$ = 2.5, the punishment fine
$\gamma$ = 5$\beta$ (see Sec. {\em Methods} for details). When
starting with an initial condition in the blue or green areas,
costly punishment can establish a behavioral norm, corresponding to
the asymptotically stable points at (0,1) and (1,0). When starting
in the red area, however, everybody in both populations will finally
behave in a non-cooperative way, as indicated by the asymptotically
stable point at (0,0). (B) Population 1 playing a stag hunt game
with model parameters \textit{B} = -1 and \textit{C} = 2, while
individuals in population 2 experience prisoner's dilemma
interactions with \textit{B} = -1 and \textit{C} = -2, assuming
equally strong populations (\textit{f} = 0.5). All individuals are
expected to end up with behavior 1, which is preferred by population
1. However, as the flow line starting with the white circle and
ending up in the asymptotically stable point (1,0) illustrates, the
evolution of the behavioral norm can take a long time and unexpected
detours. } \label{fig6}
\end{figure}

This suggests a natural tendency towards the formation of norms, whatever their content may be. Costly punishment can support the evolution of norms in prisoner's dilemma situations as well (see Fig. \ref{fig6}A). Another way of promoting the preferred coordinated behavior as commonly accepted norm is to transform a prisoner's dilemma situation into a stag hunt game in the\textit{ own} population and to make sure that the population interacts with another one with incompatible preferences and prisoner's dilemma interactions (see Fig. \ref{fig6}B). Accordingly, the sanctioning of non-conforming behavior (see Sec. {\em Methods} for details) is not the\textit{ only} mechanism to support the evolution of norms. Other cooperation-enhancing mechanisms such as kin selection (based on genetic relationship) and group selection tend to transform a prisoner's dilemma into a \textit{harmony game} (see Fig. \ref{fig5}). Therefore, genetic relatedness and group selection are \textit{not} ideal mechanisms to establish shared behavioral norms. They rather support the formation of subcultures. Moreover, the transformation of prisoner's dilemma interactions into a \textit{snowdrift game} is expected to cause polarization or conflict (see Fig. \ref{fig5}).

The evolution of language \cite{Nowak2002} is another example for the importance of norm-establishing social interactions, since successful communication requires norms, how words are used (the ``evolution of meaning'') \cite{Castello2006,Baronchelli2008}. In this connection, it is interesting to study whether the explosive development of language and culture in humans is due to their ability to transform interactions into norm-promoting stag hunt interactions. From this point of view, repeated interactions thanks to human settlements, the development of reputation mechanisms, and the invention of punishment institutions should have largely accelerated cultural evolution \cite{BoydRicherson1985,BoydRicherson1994,Gintis2009}.

Another interesting research direction relates to the circumstance that people do not only \textit{follow} norms---at the same time, they also\textit{ create} norms. This touches the issue of norm \textit{emergence} \cite{Opp1979,Opp1982,Opp2001,Horne2001a,Horne2001b,Kitts2006}. In order to address it, one also has to answer questions such as the following: How is the ``content'' of norms generated or selected, i.e. how does the prescription of a behavioral role or normative behavior come about \cite{BoydRicherson2005}? How and why do people start sanctioning non-conforming individuals, although this is costly, and why is there a tendency towards conformity at all \cite{Oliver1980}? This is beyond the scope of this paper. The same applies to related questions such as the following: Do norms always establish a ``Pareto'' or a ``system optimum'' \cite{Banfield1967,Putnam1994,Posner2000,Centola2005}? Do norms always emerge, when they would be ``functional'' or beneficial \cite{Elster1989b,Coleman1990}? Do they disappear, when they are not beneficial anymore? Do norms reduce or produce conflicts \cite{Stouffer1949}? How can one explain local conformity, global diversity, and punctuated equilibria \cite{Epstein2001}? These points will be addressed in a forthcoming publication, while the goal of \textit{this} manuscript was to develop a unified theoretical concept allowing one to study the interaction of populations with incompatible interests. This became possible by analysis of multi-population game-dynamical (replicator) equations, which have been used here to address, besides the evolution of norms, the occurrence of polarization or conflict, the outbreak of revolutions and several other relevant questions, like the importance of the power of a population and of the previous history (``initial conditions'') for the outcome of social interactions.

\section*{Methods}

\subsection*{Multi-population game-dynamical replicator equations}

Multi-population game-dynamical replicator equations \cite{Schuster1981,Helbing1992,HelbingJohansson2010} describe the temporal evolution of the proportions $p_i^a(t)$ of individuals showing behavior \textit{i} at time \textit{t} in population \textit{a}, assuming that more successful behaviors spread, as these are imitated by individuals of the same population at a rate proportional to the increase in the expected success \cite{Helbing1992,Helbing1996,Schlag1998}. The expected success is determined from the frequency of interactions between two behaviors \textit{i} and \textit{j}, and by the associated payoffs $A_{ij}^{ab}$. Focusing on the above-mentioned social dilemmas, in the case of two interacting populations \textit{a}, \textit{b} $\in$ \{1, 2\} and two behavioral strategies \textit{i}, \textit{j} $\in$ \{1, 2\}, we assume the following for interactions within the \textit{same} population \textit{a}: If two interacting individuals show the \textit{same} behavior \textit{i}, both will either receive the payoff $r_a$ or $p_a$. If we have $r_a \ne p_a$, we call the behavior with the larger payoff $r_a$ ``preferred'' or ``cooperative'', the other behavior ``non-cooperative'' or ``uncooperative''. When one individual chooses the cooperative behavior and the interaction partner is uncooperative, the first one receives the payoff $s_a$ and the second one the payoff $t_a$. To model conflicts of interests, we assume that population \textit{a} = 1 prefers behavior \textit{i} = 1 and population 2 prefers behavior 2. Therefore, if an individual of population 1 meets an individual belonging to population 2 and both show the same behavior \textit{i}~= 1, the first one will earn $R_1$ and the second one $P_2$, as behavior \textit{i} = 1 is considered uncooperative in population 2. Analogously, for \textit{i} = 2 they earn $P_1$ and $R_2$. If the interaction partners choose \textit{different} behaviors \textit{i} and \textit{j}, they earn $S_a$, when the behavior corresponds to their cooperative behavior, otherwise they earn $T_a$ \cite{HelbingJohansson2010}. In mathematical notation, the payoff matrix ($A_{ij}^{11}$) for individuals belonging to population 1 in interactions with other individuals of the same population is

\par\begin{center}
\begin{eqnarray}
\begin{array}{llcc}
&                       &   \multicolumn{2}{c}{\text{Interaction
partner's behavior}}\vspace{2mm} \\
&                       &   j=1 \text{ (preferred)}   &   j=2
\vspace{2mm} \\
\text{Focal} & i=1 \text{ (preferred)}   &   r_1           &   s_1   \\
\text{agent's} & & \\
\text{behavior} & i=2                        &   t_1           &   p_1   \\
\end{array}
\end{eqnarray}
\end{center}

while the payoff matrix ($A_{ij}^{12}$) for interactions with individuals of population 2 is

\par\begin{center}
\begin{eqnarray}
\begin{array}{llcc}
&                       &   \multicolumn{2}{c}{\text{Interaction
partner's behavior}} \vspace{2mm} \\
&                       &   j=1 & j=2   \text{ (preferred)}
\vspace{2mm} \\
\text{Focal} & i=1 \text{ (preferred)}   &   R_1           &   S_1   \\
\text{agent's} & & \\
\text{behavior} & i=2                        &   T_1           &   P_1   \\
\end{array}
\end{eqnarray}
\end{center}


To reflect incompatible preferences of both populations, we assume
that the payoff matrix of individuals belonging to population 2 is
``inverted'' or ``mirrored''. When interacting with individuals of
population 1, the related payoff matrix ($A_{ij}^{21}$) is

\par\begin{center}
\begin{eqnarray}
\begin{array}{llcc}
&                       &   \multicolumn{2}{c}{\text{Interaction
partner's behavior}} \vspace{2mm} \\
&                       & j=1 \text{ (preferred)}   &   j=2
\vspace{2mm} \\
\text{Focal} & i=1   &   P_2           &   T_2   \\
\text{agent's} & & \\
\text{behavior} & i=2    \text{ (preferred)}                 &   S_2           &   R_2   \\
\end{array}
\end{eqnarray}
\end{center}
while the payoff matrix ($A_{ij}^{22}$) when interacting with
individuals of population 2 is
\par\begin{center}
\begin{eqnarray}
\begin{array}{llcc}
&                       &   \multicolumn{2}{c}{\text{Interaction
partner's behavior}}   \vspace{2mm} \\
&                       & j=1 & j=2 \text{ (preferred)} \vspace{2mm} \\
\text{Focal} & i=1   &   p_2           &   t_2   \\
\text{agent's} & & \\
\text{behavior} & i=2        \text{ (preferred)}             &   s_2           &   r_2   \\
\end{array}
\end{eqnarray}
\end{center}

Assuming constant preferences and fixed relative population strengths $f_a$, the resulting coupled game-dynamical replicator equations for the temporal evolution of the proportion \textit{p}(\textit{t}) = $p_1^1(t)$ of cooperative individuals in population 1 and the fraction $q(t) = q_2$ of cooperative individuals in population 2 become

\begin{equation}
    \frac{dp(t)}{dt} = \begin{matrix}\underbrace{p(t)[1-p(t)]} & \underbrace{\Big[ b_1f + (c_1-b_1)fp(t) + C_1(1-f) + (B_1 - C_1)(1-f)q(t) \Big]}\\ \text{saturation factors} & \text{growth factor $F(p,q)$ containing interaction effects}\end{matrix}
\label{eq1}
\end{equation}

and

\begin{equation}
    \frac{dq(t)}{dt} = \begin{matrix}\underbrace{q(t)[1-q(t)]} & \underbrace{\Big[ b_2(1-f) + (c_2-b_2)(1-f)q(t) + C_2f + (B_2 - C_2)fp(t) \Big]}\\ \text{saturation factors} & \text{growth factor $G(p,q)$ containing interaction effects}\end{matrix}
\label{eq2}
\end{equation}

Here, we have used the abbreviation $f = f_1 = 1 - f_2$. As
indicated above, this parameter can reflect the relative population
size of population 1. More generally, however, $A_{ij}^{ab}f_b$ can
be considered to represent ``effective payoffs'', and \textit{f} can
be used to model the ``power'' of population 1 (which may not only
depend on population size, but also on education, the availability
of weapons or technologies, and other factors). $b_a = s_a - p_a$,
$B_a = S_a - P_a$, $c_a = r_a - t_a$, and $C_a = R_a - T_a$ are
payoff-dependent model parameters, which can be positive, negative,
or zero. When setting $b_a = B_a = B$ and $c_a = C_a = C$ for
simplicity, the payoff depends on the own behavior \textit{i} and
the behavior \textit{j} of the interaction partner only, but not on
the population he/she belongs to (i.e. in- and out-group
interactions just determine whether an interaction partner may be
imitated or not, but they do not influence the payoff). Given the
values of $B$ and $C$ used in our computer simulations, it is easily
possible to construct related payoff matrices. Fixing values for $P$
(e.g. $P=0$) and for $D=R-P$, we have $R(P,D)=P+D$,
$T(P,D)=R-C=P+D-C$, and $S(P,D)=P+B$.

In reality, in-group and out-group interactions may, of course, affect the payoff as well. Such situations can be treated by choosing different values for the lower-case and upper-case parameters. It is obvious that this creates a multitude of additional cases, which deserve to be investigated in detail. However, before doing so, one first has to understand the basic case addressed in this study, which is already quite complicated.

\subsection*{Specification of Costly Punishment and Effects of Group Pressure}

Let us now consider costly punishment analogously to the way it was specified in Ref.~\cite{Traulsen2009}. Then, we have $s_1 = S - \beta_1$, $t_1 = T - \gamma_1$, $S_1 = S - \beta_1 - \gamma_2$, $S_2 = S - \beta_2 - \gamma_1$, $t_2 = T - \gamma_2$, $s_2 = S - \beta_2$. This specification assumes that someone who receives the low ``sucker's payoff'' \textit{S}  in the event of unilateral cooperation (from his/her point of view), will punish the respective interaction partner immediately, which modifies the payoffs. The punishment performed by an individual of population \textit{a} reduces the payoff of his/her interaction partner by the fine $\gamma_a>0$. However, punishment is costly (it needs some punishment effort), which reduces the punisher's payoff by $\beta_a>0$. Usually one assumes $\gamma_a>\beta_a$. The correspondingly changed payoffs imply the parameters $b_1 = B - \beta_1$, $c_1 = C + \gamma_1$, $B_1 = B - \beta_1 - \gamma_2$, $C_1 = C$ and $b_2 = B - \beta_2$, $c_2 = C + \gamma_2$, $B_2 = B - \beta_2 - \gamma_1$, $C_2 = C$, which must be inserted into Eqs.~(\ref{eq1}) and (\ref{eq2}). Therefore, punishment transforms the prisoner's dilemma\textit{ within} a population into a stag hunt game, when $c_a = C + \gamma_a>0$, i.e. $\gamma_a>|C|$. The interaction with the \textit{other} population remains a prisoner's dilemma, since $B_a<0$ and $C_a<0$. Altogether, costly punishment results in the modified two-population game-dynamical equations

\begin{equation}
    \frac{dp(t)}{dt} = p(t)[1-p(t)] \Big[ F(p,q) - \beta_1 f[1-p(t)] + \gamma_1fp(t) - (\beta_1 + \gamma_2)(1-f)q(t) \Big]
\label{eq3}
\end{equation}

and

\begin{equation}
    \frac{dq(t)}{dt} = q(t)[1-q(t)] \Big[ G(p,q) - \beta_2 (1-f)[1-q(t)] + \gamma_2(1-f)q(t) - (\beta_2 + \gamma_1)fp(t) \Big]
\label{eq4}
\end{equation}

These can generate $dp/dt \ge 0$ and $dq/dt \ge 0$ even for the prisoner's dilemma with \textit{B} $<$ 0 and \textit{C} $<$ 0. While punishment in a \textit{one}-population prisoner's dilemma can lead to \textit{cooperation} \cite{Traulsen2009}, in the \textit{multi}-population case it can cause \textit{normative behavior} (see green and blue areas in Fig. \ref{fig6}A).

Rather than considering costly punishment as discussed before, one may also consider that individuals can apply \textit{ group pressure} to support conformity and discourage discoordinated behavior \cite{CialdiniTrost1998,FehrGachter2002}. That could be reflected by subtracting a value \textit{$\delta $} from the off-diagonal payoffs \textit{S} and \textit{T} or by adding \textit{$\delta $} to the diagonal elements \textit{R} and \textit{P}. This results in the effective model parameters $b_a = B_a = B - \delta$ and $c_a = C_a = C + \delta$ \cite{HelbingLozano2010}. Therefore, if the group pressure \textit{$\delta $} is large enough (namely, $\delta>$ \textbar \textit{C}\textbar ), a prisoner's dilemma with \textit{B} $<$ 0 and \textit{C} $<$ 0 is transformed into a stag hunt game with $b_a = B_a<0$ and $c_a = C_a>0$.

\subsection*{Discussion of Implicit Model Assumptions}

Any model has some underlying model assumptions, and even though they may not be exactly fulfilled, the resulting model can be a useful approximation. The evolutionary game theoretical model of this paper assumes that individuals show a certain behavior and stick to it until they change it due to social learning (e.g. imitation) or mutations (e.g. trial-and-error behavior). This appears applicable to situations of routine choice \cite{Gintis2009} and to situations, where individuals do not spend much time on analyzing situations, but rather orient at what the others do. Crowd behavior and certain kinds of public opinion formation seem to be good examples for this \cite{Helbing1992}. The approach should also be applicable to many kinds of culturally acquired behaviors, including a considerable number of behavioral conventions and norms.

In contrast to evolutionary game theory, classical game theory assumes complex, strategic decision-making processes based on utility maximization. These decision-making processes usually consider individual preferences of interaction partners and other aspects. In spite of this difference, both approaches lead to mutually consistent outcomes in many cases. For this reason, it should not matter for our main conclusions whether the analysis is based on classical or evolutionary game theory. Indeed both kinds of game-theoretical analysis are expected to show the four archetypical types of system behaviors identified in our paper. In this connection, it is interesting to note the following \cite{Gintis2000}:

\begin{enumerate}
\item  Every Nash equilibrium is a fixed point of the replicator dynamics and the game-dynamical equation. (At a Nash equilibrium, which may also correspond to a mixed strategy, no player can improve the payoff by changing the strategy unilaterally.)

\item  Every (asymptotically or neutrally) stable fix point of the replicator equation is a Nash equilibrium.

\item  The fix points (and their stability properties) imply the main conclusions of our paper, as they determine the features of the system dynamics, which are reflected by the basins of attraction and the flow lines.
\end{enumerate}

Further assumptions underlying the multi-population evolutionary game-theoretical model become obvious in the mathematical derivation of the equations. One- or multi-population replicator equations may describe the spreading of more successful individuals via a higher reproduction rate (see, e.g. \cite{HofbauerSigmund1998}. However, they may also reflect social learning, namely by the imitation of more successful behaviors \cite{Helbing1992,Helbing1996,HelbingJohansson2010}. The above model equations result in case of proportional imitation \cite{Helbing1992,Schlag1998}. They assume that interactions take place in large, well-mixed populations, usually between different individuals. If effects of repeated interactions (and, thereby, a ``shadow of the future'') shall be taken into account, this can be done by modifying the payoffs accordingly \cite{Nowak2006,HelbingLozano2010}. A similar thing applies to reputation effects, network reciprocity, group selection, etc. (see also Fig. \ref{fig5}). It is furthermore possible to generalize the approach to finite populations \cite{Taylor2004} and to populations with spatial or network interactions (see, for example, \cite{Roca2009,SzaboFath2007,FlacheHegselmann2001}.

Entities belonging to different populations differ in two aspects: They earn different payoffs, and they imitate different entities (namely, better-performing entities belonging to the \textit{own} population, assuming that it would not necessarily be wise to copy behaviors of individuals with different preferences and payoff functions). If, additionally, in-group interactions shall be distinguished from out-group interactions (in the sense that not only the \textit{actual} behavior, but also the \textit{preferred} behavior of the interaction partner matters), one has to specify the parameters $b_a$ and $c_a$ differently from the parameters $B_a$ and $C_a$. In this way, one can even consider cases, where both populations play different games (see Fig. \ref{fig6}B).


\section*{Acknowledgments}

The authors 
are grateful to Thomas Chadefaux, Andreas Flache, Ryan Murphy, Carlos Roca, Stefan Bechtold, Sergi Lozano, Heiko Rauhut, Wenjian Yu, and further colleagues for valuable comments and to Sergi Lozano for drawing Fig. \ref{fig5}. They also like to acknowledge concrete suggestions of the anonymous referees regarding possible improvements of the manuscript. Furthermore, D.H. thanks Thomas Voss for his insightful seminar on social norms.


\clearpage


\end{document}